\documentclass[epjST]{svjour}
\usepackage[latin10]{inputenc}

\usepackage{bm}        % for math

\usepackage{graphics}
\usepackage{amsfonts}
\usepackage{amssymb}
\usepackage{amsmath}
\usepackage{textgreek} 
\usepackage{float}

\usepackage[T1]{fontenc}
\usepackage{dcolumn}   % needed for some tables
\usepackage{bm}        % for math

\begin{document}
%%%%%%%%%%%%%%%%%%%%%%%%%%%%%%%%%%%%%%%%%%%%%%%%%%%%%%%%%%%%%%%%%%%%%
\title{Very long-range attractive and repulsive forces in Model Colloidal Dispersions}
\author{Alfredo Gonz\'{a}lez-Calder\'{o}n\inst{1}\fnmsep\thanks{Present address: CONACyT - Centro de Ingenier\'ia y Desarrollo Industrial (CIDESI), Av. Playa Pie de la Cuesta 702, Desarrollo San Pablo, 76125 Quer\'etaro, Qro., M\'exico. \email{alfredo.gonzalez@cidesi.edu.mx}} \and Enrique Gonz\'alez-Tovar\inst{2}\fnmsep\thanks{\email{henry@mail.ifisica.uaslp.mx}} \and Marcelo Lozada-Cassou\inst{1}\fnmsep\thanks{Corresponding author. \email{marcelolcmx@ier.unam.mx}}}
\institute{Instituto de Energ\'ias Renovables, Universidad Nacional Aut\'onoma de M\'exico (U.N.A.M.), 62580 Temixco, Mor., M\'exico. \and Instituto de F\'isica, Universidad Aut\'onoma de San Luis Potos\'i, \'Alvaro Obreg\'on 64,
78000 San Luis Potos\'i, S.L.P., M\'exico.}
%
%%%%%%%%%%%%%%%%%%%%%%%%%%%%%%%%%%%%%%%%%%%%%%%%%%%%%%%%%%%%%%%%%%%%%%
\abstract{
Experiments with polymer latex solutions show the coexistence of order-disorder structures of macroions. Because of the large macroions' sizes, this order-disorder phase coexistence imply the existence of very long-range attractive and repulsive forces, which can not be explained in terms of conventional direct interaction potentials, which are short-range. Here we apply an integral equations theory to a simple model for colloidal dispersions, at finite concentrations, calculate the particles distribution functions and the involved effective forces. We find very long-range attractive and repulsive forces among the like-charged macroions. The distribution functions are in qualitative agreement with experimental results. The origin of these forces are discussed in terms of an energy-entropy balance.
}
\maketitle

%%%%%%%%%%%%%%%%%%%%%%%%%%%%%%%%%%%%%%%%%%%%%%%%%%%%%%%%%%%%%%%%%%%%%%

\section{Introduction}
\label{sec:1}
The study of the interactions of charged colloidal particles or macromolecules is relevant in the fields of physics, chemistry, biology, energy and technology~\cite{pieransky83,kamenetzky94,wennerstrom,vatamanu}. In particular, suspensions of highly charged colloidal particles or macromolecules, depending on the volume fractions, salt concentrations, and temperature, can form gas, liquid and crystal structures~\cite{clark1979,ise80,ottewill1989}, and order-disorder coexistence regions~\cite{ise83,ise1993,ise94,tata1997}. Because of the large macroions' sizes, this experimentally found order-disorder phases coexistence imply the existence of very long-range repulsive and attractive forces, which can not be explained in terms of conventional direct interaction potentials, which are of relatively short range~\cite{israelachviliNature08}. Particularly puzzling is the existence of a long-range attraction among like-charged particles. To explain this long-range attraction, semi-phenomenological long-range potentials were proposed~\cite{sogami84,ise96,sogami2012}.  Both, the theoretical approach and the experimental results have been questioned by different authors~\cite{palberg1994,grier1996,hansen97,schmitz02}, either arguing the semi-phenological nature of the proposed potential and/or that the experimentally observed ordered-disorder coexistence was a result of the system not being in equilibrium and/or being in confinement. However, the criticisms to the experimental results seem to have been properly responded~\cite{tata1994,tata1997,ise2000,tata2008}. Hence, the main experimental finding, i.e., \textit{there is a long-range attraction between like-charged particles, at finite colloids concentration}, is probably correct. But, then, does this attractive force exists, and if it does exists what is the origin of this long-range attraction? N. Ise et al.~\cite{ise66,sogami84,ise2010} have proposed that this attraction is due to ``\textit{an intermacroion attraction through the intermediary of counterions balanced by intermacroion repulsion}''. While in this article we do not try to find an order-disorder coexistence curve, we do address the above questions, and calculate the macroion-macroion forces, through a well established liquid theory. We do, indeed, find a long-range attraction force. We hope to help to clarify this controversy, and offer a mechanism to explain this attraction. In this paper we extend our previous calculations for macroions dispersions, at finite concentrations~\cite{manzanilla2013}, where we studied the effect of the location of the macroions charge, and reported a long-range macroion-macroion correlation. Here we provide new results for macroions forces, give a detailed analysis of their foundations, nature and range, offer an explanation for our reported long-range attractive-repulsive forces, and exhibit a comparison with existing experimental data. In Section~\ref{sec:2} we present our theoretical approach. In Subsection~\ref{sec:3} we outline the derivation of the integral equations. In Subsection~\ref{sec:4} we describe the model for a macroions solution, at finite volume fraction, and outline the formulation of the electrostatic and forces in the system. In Section~\ref{sec:5} we report the results for the radial distribution functions, the electrical field, and the mean forces among the like-charged macroions. Finaly, in Section~\ref{sec:6} we discuss our results and give some conclusions.

%%%%%%%%%%%%%%%%%%%%%%%%%%%%%%%%%%%%%%%%%%%%%%%%%%%%%%%%%%%%%%%%%%%%%
\section{Theory}
\label{sec:2}

Most studies on charged colloidal suspensions are made for isolated macroions or two like-charged colloids particles, i.e., at infinite dilution, immersed into a model electrolyte~\cite{verwey48,stillinger1960,mcquarrie80,schmitz93,Kjellander86,levin99,belloni2000,levin2016}. Theoretical and simulation investigations of nano-particles suspensions at finite volume fractions are technically more difficult due to the large increase of integration space (with different length scales) in theoretical equations, or the huge number of particles in, necessarily, larger simulation boxes. Thus, many studies used some kind of charge renormalization~\cite{belloni1998,bocquet2002,ise99,medinanoyola2011}, colloidal suspensions with no salt added~\cite{linse99}, or very low salt concentration~\cite{hansen97}. In general these investigations are for relatively low colloidal charge and volume fraction~\cite{linse00,lobaskin03}. 

Since the pioneer development of the integral equations by Kirkwood, et al.~\cite{poirier1954,stillinger1960}, for homogeneous and inhomogeneous size-symmetrical electrolytes, based on a density expansion plus a superposition approximation, other approximated integral equations based on the Ornstein-Zernike equation~\cite{McQuarrie} have been derived. Among them is the Hypernetted-Chain/Mean Spherical Approximation (HNC/MSA)~\cite{Lozada-Cassou92a,attard96}, first proposed by S. L. Carnie et al., in 1981~\cite{carnie1981}. In this report we solve these integral equations for a model of a three species electrolyte solution. In this section we outline the derivation of the HNC/MSA equation, with which we will obtain the radial distribution functions of the macroions solution, describe the macroions' model, give the expressions of the relevant electrostatic variables and interparticle's mean forces, in terms of the radial distribution function.  

%%%%%%%%%%%%%%%%%%%%%%%%%%%%%%%%%%%%%%%%%%%%%%%%%%%%%%%%%%%%%%%%%%%%
\subsection{The Hypernetted-Chain/Mean Spherical equations }
\label{sec:3}

In our model the particles are taken to be charged, hard spheres, with different charges and diameters, such that one of the species is much larger than the other two, in order to model a macroions dispersion at finite concentration. We will refer to his model as the Colloidal Primitive Model (CPM)~\cite{manzanilla2011a,manzanilla2011b,manzanilla2013}, which is a variant of the well known primitive model~\cite{kjellander1992,zuckerman2001}. 

The multicomponent Ornstein-Zernike (OZ) equation for a fluid of $n$ species is given by $h_{i j}({\bf r}_{21}) = c_{ij}(\mathbf{r}_{21} ) + \sum_{l=1}^{n}\rho_{l} \int_{V} h_{i l}({\bf r}_{23})
c_{lj}(\mathbf{r}_{31} )d\mathbf{r}_{3}$, where $V$ is the system's volume,  $i,j =1,2,\dots,n$, and $\rho_l$ is the number density, of species $l$. $h_{ij}(\mathbf{r}_{21}) \equiv g_{ij}(\mathbf{r}_{22})-1$ is the total correlation functions for two particles at
 $\mathbf{r}_{2}$ and $\mathbf{r}_{1}$, of species $i$ and $j$, respectively,
with  $\mathbf{r}_{21}= \mathbf{r}_{2}- \mathbf{r}_{1}$, such that $g_{ij}({\bf r}_{21})$ is the pair correlation function, also referred to as the {\em radial distribution function (RDF)}, which gives the probability density of finding a particle $1$, of species $j$, at the distance $\mathbf{r}_{21}$, from the central particle $2$, of species $i$. The OZ equation is a probability balance, that guaranties a constant chemical potential in the whole system~\cite{odriozola2017}. To solve it, closure relations for the direct correlation functions, $c_{ij}(\mathbf{r}_{21} )$ and $c_{lj}(\mathbf{r}_{31})$, are needed. These function are basically approximations for a quasi-particle, in the context of many-body theory~\cite{Friedmanbook}. Among others, two direct correlation approximations have been successfully used, the Hypernetted-Chain (HNC), defined as $c_{ij}(\mathbf{r}_{21}) = -\beta u_{ij}(\mathbf{r}_{21}) +h_{il}(\mathbf{r}_{21})-\ln{g_{i j}({\bf r}_{21}) }$, and the Mean Spherical (MSA) as $c_{il}(\mathbf{r}_{21}) = -\beta u_{ij}(\mathbf{r}_{21})$, where $\beta=1/(kT)$, $k$ is the Boltzmann constant, $T$ is the system temperature, and $ u_{ij}(\mathbf{r}_{21})$ is the, unscreened, pair-interaction potential between particles 1 and 2. If the HNC approximation is used in the first term of the right-hand side of the OZ equation, we obtain

\begin{eqnarray}
\begin{split}
g_{i j}({\bf r}_{21}) = \exp \left\{-\beta u_{ij}(\mathbf{r}_{21}) +
  \sum_{l=1}^{n}\rho_{l} \int_{V} h_{i l}({\bf r}_{23})
c_{lj}(\mathbf{r}_{31} )d\mathbf{r}_{3} \right\}. \\
\label{trainee}
\end{split}
\end{eqnarray}

\noindent If $c_{lj}(\mathbf{r}_{31})$, inside the integral in Eqs.~\eqref{trainee}, is taken to be given by the MSA, for which there is an analytical solution for charged fluids~\cite{kazuo77}, we get the HNC/MSA integral equations, for an homogeneous fluid of $n$-species. The use of the MSA allows to have analytical kernels in Eq.~\eqref{trainee}. A detail account of the derivation of Eq.~\eqref{trainee} for the CPM is given in Manzanilla-Granados et al.~\cite{manzanilla2013}. 

%%%%%%%%%%%%%%%%%%%%%%%%%%%%%%%%%%%%%%%%%%%%%%%%%%%%%%%%%%%%%%%%%%%%
\subsection{The model, radial distribution functions, electrostatics and forces }
\label{sec:4}

For a system of three species fluid, equations~\eqref{trainee} become a set of seven, coupled, non-linear integral equations, which because we are taking advantage of the analytical solution for the MSA can be reduced to three coupled integral equations. We numerically solve these equations with a finite elements method developed in the past for the solution of integral equations~\cite{sanchez92,Yeomans1,manzanilla2013}.

We solve Eqs.~\eqref{trainee} for the CPM, for a three species charged fluid, i.e., positive ions, negative ions and macroions, of species $+, -$, and $M$, respectively. For simplicity, we will assume the macroions to be positive large particles of diameter $a_{M}$, and all the little ions of equal size, including the macroions' counterions (the case in which the macroions are negatively charged is symmetrical to the positively charged macroions case). Thus, the ions's diameters are $a_{+}=a_{-}\equiv a$. Therefore, from the solution of Eqs.~\ref{trainee} we obtain the macroion-macroion, $g_{MM}(r)$, macroion-anion, $g_{M-}(r)$, and macroion-cation, $g_{M+}(r)$, radial distribution functions. The other three distribution functions, i.e., $g_{- -}(r)$, $g_{+ +}(r)$, and $g_{+ -}(r)$ are analytically obtained from the MSA~\cite{kazuo77}. $g_{- +}(r)$=$g_{+ -}(r)$, since we are assuming equal size for the little ions.

\noindent The interaction potential between two particles of species $i$  and $j$, with a separation distance $r$, is given by
 \begin{equation}
 u_{i j}(r)=
 \left \{ \begin{array}{ll}
 \infty &\text{for $r< a_{ij}$}\\
 {\displaystyle\frac{e^{2}z_{i}z_{j}}{\epsilon r}}& \text{for $r \geq a_{ij}$} 
 \end{array} \right.  \text{with } i,j=+,-,M.
 \label{mar1}
 \end{equation}
\noindent where $a_{ij} = (a_{i} + a_{j})/2$, $e$ is the proton charge, $z_i$ and $z_j$ are the valances of particles of species $i$ and $j$, respectively, and $\epsilon$ is the solvent dielectric constant, and such that the system is electroneutral ($\sum_{j=+,-,M} ez_{j}\rho_{j}=0$). Both the electrolyte ions and macroions are assumed to have the same dielectric constant as that of the solvent, to avoid image charges. For a colloid's valence $Z_M$, its surface charge density is $\sigma_M = Z_M e/(\pi a_M^2)$. Hence, its unscreened charge is $Q_M=\pi a_M^2\sigma_M$. However, in general, the colloid's effective charge, at a distance $r$ from its center (i.e., its bare charge $Q_M$, plus the induced charge in the surrounding fluid, up to a distance $r$), is given by 

\begin{eqnarray}
Q_M(r)=Q_M+4\pi \int_{a_M/2}^{r}\rho_{el}(t)t^{2}dt=- 4\pi \int_{r}^{\infty}\rho_{el}(t)t^{2}dt
\label{charge}
\end{eqnarray}
 
\noindent where

\begin{eqnarray}
\rho_{el}(r) \equiv \sum_{j=+,-,M} ez_{j}\rho_{j}g_{M j}(r)
\label{reloj2}
\end{eqnarray}

\noindent is the charge concentration profile around the central colloidal particle, and we have used the electroneutrality condition for the bulk fluid, given by 

\begin{eqnarray}
Q_M =- 4\pi \int_{a_M/2}^{\infty}\rho_{el}(t)t^{2}dt.
\label{neutrality}
\end{eqnarray}

\noindent Eq.~\eqref{charge} implies that as $r\rightarrow\infty$, $Q_M(r) \rightarrow 0$. While $Q_M$ and $Q_M(r)$ gives the macroions bare charge, and  effective charge at every distance $r$ from the center of the reference macroion, the surface charge density is often used in the literature to report colloids' experimental and theoretical studies. Hence, we define the charge density profile, $\sigma_{M}(r)$, around the central macroion, as

\begin{eqnarray}
\sigma_{M}(r) \equiv \sigma_{M}\frac{R_{M}^2}{r^2} +
\frac{1}{r^2}\int_{R_{M}}^{r} \rho_{el}(t)t^{2}dt,
\label{twoa}
\end{eqnarray}

\noindent where $R_M=a_M/2$ is the macroion radius, and $\sigma_{M}$ is its surface charge density, and the second term in Eq.\eqref{twoa} is the induced charge in the fluid around the central particle from $R_{M}$ to $r$.  In terms $\sigma_M$, Eq.~\eqref{neutrality} can be written as

\begin{eqnarray}
\sigma_{M} = - \frac{1}{R_{M}^2}\int_{R_{M}}^{\infty} \rho_{el}(t)t^{2}dt.
\label{two-bis}
\end{eqnarray}

\noindent 

 Hence, Eq. \eqref{twoa} can also be expressed as

\begin{eqnarray}
\sigma_{M}(r) = - \frac{1}{r^2}\int_{r}^{\infty} \rho_{el}(t)t^{2}dt.
\label{two}
\end{eqnarray}

\noindent By Gauss' law, the electrical field, $E_{M}(r)$, at the distance $r$ from the center of the reference particle, is $E_{M}(r)= 4\pi\sigma_{M}(r)/\epsilon$.

On the other hand, Eqs.~\eqref{trainee} can be recast as~\cite{manzanilla2013} 

\begin{eqnarray}
g_{i j}({\bf r}_{21}) = \exp\left\{-\beta q_j\psi_{ij}({\bf r}_{21}) + \beta\omega^{sr}_{ij}({\bf r}_{21})\right\},
\label{trainee2}
\end{eqnarray}

\noindent where $\psi_{ij}({\bf r}_{21})$, is the mean electrostatic potential, given by

\begin{eqnarray}
\psi(r) = \frac{4 \pi}{\epsilon}\int_{r}^{\infty}
\rho_{el}(t)
\left(t-\frac{t^2}{r} \right)dt,
\label{reloj}
\end{eqnarray}

\noindent and $ \omega^{sr}_{ij}({\bf r}_{21})$  is a mean potential of short-range force. Thus, the potential of mean
force, given by

\begin{equation}
W_{ij}({\bf r}_{21}) \equiv -kT\ln{g_{ij}(r)} =ez_{j} \psi_{ij}({\bf r}_{21})+\omega^{sr}_{ij}({\bf r}_{21}),
\label{PMF}
\end{equation}

\noindent contains the contributions from the electrostatic and short-range correlations, between the central particle, of species $i$, and another particle in the fluid, of species $j$. Therefore, the total mean force is equal to the sum of the electrostatic and short-range components, i.e.,

\begin{equation}
 F^{T}_{ij} (r)= F^{e}_{ij}(r) + F^{s}_{ij}(r),
\label{madru1}
\end{equation}

\noindent with $F^{T}_{ij}({ r}) = - d(W_{ij}({r}))/dr$, $F^{e}_{ij} = -ez_{j} d(\psi_{ij}({r}))/dr= ez_{j}E_{M}(r)$ and $ F^{s}_{ij}= -d(\omega^{sr}_{ij}({r}))/dr $ respectively, where we have taken advantage of the spherical symmetry of the system, and have omitted the subindexes for simplicity. Here, we will refer to the short-range component of the force as the entropic component.

\textit{Although the mean force is equal to the sum of the entropic and electrostatic force components, these two components are entangled or intertwined, through the non-linear integral Eq.~\eqref{trainee2}}, since both components (electrostatic and entropic) are dependent functionals of the RDF, $g_{ij}(r)$. The entanglement of the entropy and energy contributions to the total system energy is beyond the non-linearity of Eq.~\eqref{trainee2}, as can be seen from Eq.~\eqref{PMF}; i.e., a change in $Q_M$, $\rho_i$ and $a_i$, for any of the ionic species, $\epsilon$ and/or $T$ will modify \textit{simultaneously} $\psi_{ij}({\bf r}_{21})$ and $ \omega^{sr}_{ij}({\bf r}_{21})$. This entanglement of the electrical and entropic components of the fluid energy should not be confused with its use in quantum mechanics or mathematical psychology, among other science fields~\cite{lawless}.

The total force between the central macroion, of species $M$, and an ion of species $i$, $F^{T}_{Mi}(r)$, was calculated by numerical derivation of the potential of mean force, i.e., $F^{T}_{Mi}(r)=-dW_{Mi}(r)/dr=kTdln[g_{Mi}(r)]/dr$. Its electrostatic component, $F^{e}_{Mi}(r)$, was calculated by numerical derivation of Eq.~\eqref{reloj} and/or directly from $eZ_{M}E_{M}(r)= 4\pi eZ_{M}\sigma_{M}(r)/\epsilon$, with the aid of Eq.~\eqref{two}. Its entropic component, $F^{s}_{Mi}(r)$,  was calculated from the difference of the total and electrical forces, and/or through the contact values of the different radial distribution function, $g_{Mi}(r=a_{Mi}/2)$, i.e., with $F^{s}_{Mi}(r)= \sum_{j=+,-,M} \rho_{j}g_{M j}(r;a_{Mi}/2)$. The contact values, of course, are different for different values of the position $r$ of the particle of species $i$. \textit{All the different procedures for the calculations of the total force components were found to be consistent}.

%%%%%%%%%%%%%%%%%%%%%%%%%%%%%%%%%%%%%%%%%%%%%%%%%%%%%%%%%%%%%%%%%%%%

\section{Results}
\label{sec:5}

With the radial distribution functions obtained from Eqs.~\eqref{trainee} we calculate the total force, and its electric and entropic components, for macroions at several colloidal volume fractions, $ \phi\equiv\frac{1}{6}\pi\rho_Ma_M^3$, and surface charge densities $\sigma_M$, i.e., several macroions' valences, $Z_M$. In all cases, unless otherwise indicated, the added salt is a 1:1, $0.1$M electrolyte, with ionic diameter, $a=4.25\,\mbox{\AA}$, and the macroions' diameter is $a_M=10a$. The solvent dielectric constant is taken to be $\epsilon=78.5$, and the system temperature $T=298șK$.

\subsection{Long-range attractive and repulsive forces: Detailed analysis and charge dependence.}
\label{sec:6}

\begin{figure}
\resizebox{0.9\columnwidth}{!}{
  \includegraphics{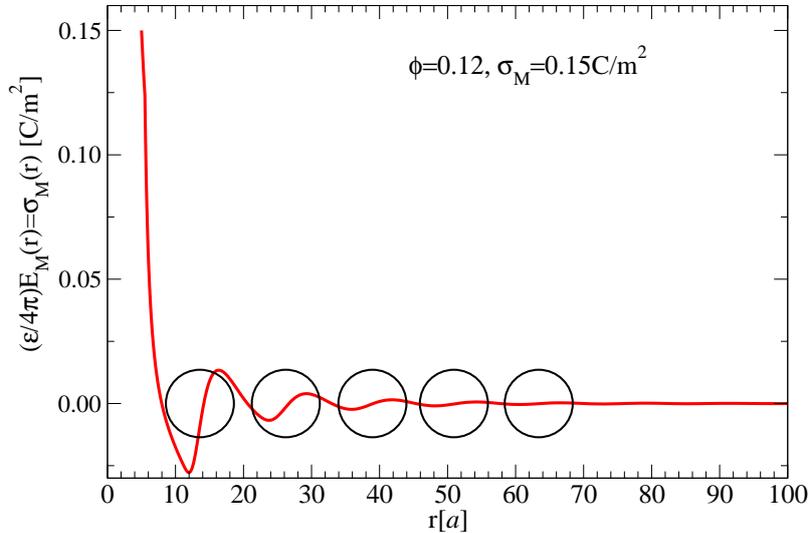} }
\caption{Electrical field and charge density profiles, around the reference macroion, measured from its center. The circles positions , in $r$, coincide with the maxima in the radial distribution function. $\rho_{salt}=0.1$M, $a_M=10a$. }
\label{Fig_1}       % Give a unique label
\end{figure}

In Fig.~\ref{Fig_1} we present the HNC/MSA results for the macroion's charge density profile, $\sigma_{M}(r)=(\epsilon/4\pi)E_{M}(r)$, as a function of the distance, $r$, to the center of the central particle, for $\phi$=0.12, and surface charge density, $\sigma_M = 0.15C/m^2 (Z_M\approx53)$, thus (of curse), from Eq.~\ref{two}, $\sigma_M(r=5a)=0.15C/m^2$, on the surface of the central, reference particle. As was pointed out above, the $\lim_{r\to\infty} \sigma_M(r)\to 0$, to satisfy the electroneutrality of the bulk fluid, and was numerically corroborated in our calculations.

A number of sequential charge inversions is observed, indicating the change of direction of the electrical field, $E_M(r)$. The locations of the maxima and minima of the charge density in the fluid are closely related to the macroions' size, volume fraction, surface charge, and location, and show interesting new surface phenomena~\cite{alfredo2017,alfredo2018}. However, in this article we addresses only to the foundations, nature and range of the forces among charged macroions.  

The dimension and location of the circles shown in Fig.~\ref{Fig_1} correspond exactly to the size and most probable location of the layers of neighboring macroions, around the reference macroion, according to the corresponding RDF (see Table~\ref{tab:1}). Their location do not  coincide with the inflection points of the charge profile curves, since these are the points of zero electrical field, but not of zero force. We will come back to this point later. Notice that the charge inversions occur before (negative), and after (positive) the position of the positively-charged macroions, indicating the effect of the adsorbed negative ions. Because the negative ions are the sum of the salt anions plus the macroions' counterions, the molar concentration of the anions, for a macroions dispersion with $\phi=0.12$, and $ \sigma_M=0.15C/m^2$, is $\rho_-=0.3634M$, which is more than three times that of the little cations ($\rho_+=0.1M$). From $r=5a$, i.e., from the surface of the central macroion, up to $r\approx12a$ the charge density is seen to rapidly decrease down to $\sigma_M(r=11.88a)\approx-0.028C/m^2)$. The position of this first correlated macroions' layer is $r\approx13.59a=1.359a_M$. Then the charge density profile increases up to $\sigma\approx0.0135C/m^2$, at $r=16.5a$. This qualitative behavior of the charge density profile continues up to around $250a$. The $\lim_{r\to\infty} \sigma_M(r)\to 0^-$, which implies that the net electrical field is, overall, cohesive, but counterbalanced by the entropic repulsion. In Table~\ref{tab:1} the distance to the central macroion of this first correlated macroions' layer increases, for very low surface charge densities, and decreases for low to high surface charge densities. We will come back to this point later in this article.

\begin{figure}
\resizebox{0.9\columnwidth}{!}{
  \includegraphics{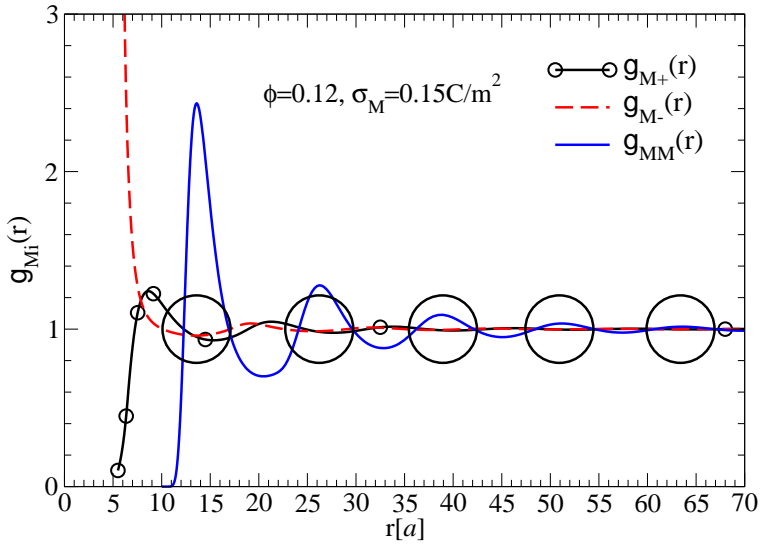} }
\caption{Radial distribution function, around the central macroion, measured from its center. The circles have the same meaning as in Fig.~\ref{Fig_1}. $\rho_{salt}=0.1$M, $a_M=10a$.}
\label{Fig_2}       % Give a unique label
\end{figure}

In Fig.~\ref{Fig_2} we show radial distribution functions for the macroions, $g_{MM}(r)$, little cations, $g_{M+}(r)$, and anions, $g_{M-}(r)$, as a function of the distance $r$, to the center of the reference macroion, and for the same solution parameters, as in Fig.~\ref{Fig_1}. The anions are strongly adsorbed to the central macroion, while the cations and macroions are repelled to some distance away, i.e., their RDFs are below their bulk values ($g_{MM}(r)$ and $g_{M+}(r)$ are below $1$). The RDF's are oscillatory, with the first six maxima of $g_{MM}(r)$ located at $r/a\approx13.59, 26.22, 38.95, 50.93, 63.40, 75.88$ (see Table~\ref{tab:1}). Hence, the wavelength of the oscillation of the macroions layers is first $\lambda \approx13.59a$, and later $\lambda\approx12.5a$. Between the second ($r\approx13.59a$) and third ($r=26.22a$) layers of macroions, the anions and cations are seen to oscillate around their bulk values, but with a relatively high adsorption of anions next to the first layer of macroions, at $r\approx13.59a$, and cations to the second layer of macroions, at $r\approx26.22a$. 

Therefore,  intermediated to two neighboring layers of macroions, can be fitted between 2 to 3.5 layers of ions. This behavior is repeated for the rest of the macroions layers. From $r\approx19a$ to $r\approx22a$, \textit{both, $g_{M-}(r)$ and $g_{M+}(r)$ are simultaneously above their bulk values}. This is a very unusual behavior. This effect is not observed in bulk electrolytes~\cite{henderson1983}, electrolytes next to a charged electrode~\cite{LozadaJCP82,gonzalez85,gonzalez89}, confined electrolytes~\cite{LozadaJCP90a} or macroions~\cite{jimenez04,odriozola2009}. An exception was reported for the interface of a cylindrical electrode, with surface charge of $0.1544 C/m^2$, and a symmetrical 1:1 electrolyte, with a relatively large ionic diameter of $7.4$\AA~\cite{gonzalez85}, showing the relevance of the ions excluded volume. However, a previous comparison of this theory with molecular dynamics calculation shows a very good qualitative, and quantitative agreement in this feature effect~\cite{manzanilla2011b}, and clearly is a consequence of the finite concentration of the large colloidal particles, i.e., excluded volume effect.

Whence, the structure around the central macroion seems to be a sequence of layers macroion-anion-cation-macroion. However, as pointed out before, for these conditions of $\phi$ and $\sigma_M$, $\rho_-=3.63\rho_+$, and thence the number of anions greatly overcomes the number of cations, around the macroions. This, suggest that the macroion-macroion attraction is, at least, partially due to the anions adsorption. 

\begin{table}
\caption{Location of the first six maxima in the corresponding RDFs, given in units of [a], measured from the center of the reference, central macroion, for six different macroions charge densities, and at a fixed volume fraction, $\phi=0.12$. Notice that because $a_M=10a$, by dividing by 10, all the distances reported in this table straightforwardly can be expressed in terms of $a_M$, a unit of measure used in several experimental reports of the structure of polystyrene latex particles.}
\label{tab:1}       % Give a unique label
\begin{tabular}{lllllll}
\hline\noalign{\smallskip}
$\sigma_M [C/m^2]$  & first & second & third &fourth &fifth & sixth \\
\noalign{\smallskip}\hline\noalign{\smallskip}
0.000& 10.00 & 24.22 & 36.45 & 48.43 &   ------ & ------ \\
0.005& 10.00 & 24.47 & 36.95 & 48.93 &   ------ & ------ \\
0.025 & 13.59 & 28.22& 42.44 & 56.91 & 70.80 & ------ \\
0.050& 14.14 & 28.47 & 42.44 & 56.42 & 70.39 & 84.87 \\
0.150& 13.59 & 26.22 & 38.95 & 50.93 & 63.40 & 75.88 \\
0.300& 12.47 & 24.72 & 36.45 & 47.93 & 59.41 & 70.89 \\
\noalign{\smallskip}\hline
\end{tabular}
\end{table}

\begin{figure}
\resizebox{0.9\columnwidth}{!}{
  \includegraphics{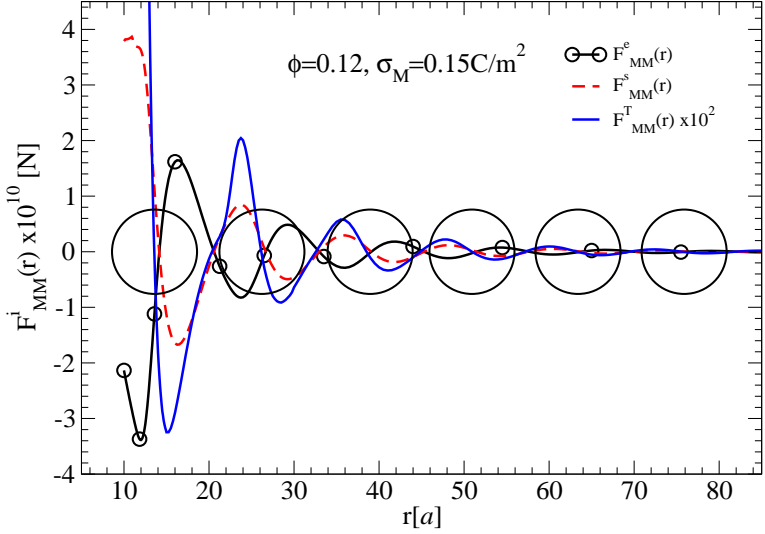} }
\caption{Mean force among like-charged macroions, and its electrical and entropic components, as a function of the distance to the center of the reference macroion. The circles have the same meaning as in Fig.~\ref{Fig_1}. $\rho_{salt}=0.1$M, $a_M=10a$.}
\label{Fig_3}       % Give a unique label
\end{figure}

In Fig.~\ref{Fig_3} the total mean force, $F^{T}_{MM}(r)$, and its electric, $F^{e}_{MM}(r)$, and entropic, $F^{s}_{MM}(r)$, components, among the positively charged macroions, are plotted as a function of the distance $r$ from the central macroion. The solution parameters are the same as in Fig.~\ref{Fig_1}. Counterintuitively the electric force is attractive, while the entropic force is repulsive, before the first layer of macroions, located at $r\approx13.59$, to become both, later, oscillatory. A comparison of Figs.~\ref{Fig_1}, \ref{Fig_2}, and \ref{Fig_3} suggest that the first electric attraction is due to the anions adsorbed to the macroions, and the later repulsive electric force is due to the charge of the first layer of macroions. On the other hand the little negative and positive ions present before the first layer of macroions produces the entropic repulsive force, as an excluded volume effect. The entropic force always overcomes the electric component. Notice that $F^{T}_{MM}(r)$ is two orders of magnitude lower than its components, so in Fig~\ref{Fig_3} we have rescaled $F^{T}_{MM}(r)$. We have drawn six circles, which have the same size as the macroions, and their locations were obtained from the corresponding RDF (see Fig.~\ref{Fig_2} and Table~\ref{tab:1}). The position of the circles correspond to the six first induced layers of macroions, around the central macroion. \textit{Their centers coincide with the points of zero total force}, as they should have, which, on the other hand, corroborates the correctness of our forces calculations.

Let us see this in some more detail. The first minimum of the electrostatic force occurs at $r\approx12a$, which corresponds to a distance around $R+7a$, from the center of the reference macroion. The reduced concentration profile of anions, i.e., $g_{M-}(r)$ (see Fig.~\ref{Fig_2}), shows a very high adsorption of negative charge, from $r=5.5a$ down to its bulk value, $g_{M-}(r=10.083a)\lesssim1.0$. Thus overscreening the central macroion charge. At $r=10a$ we see $F^{e}_{MM}(r)$ to begin its attractive behavior, and then increasing the attraction intensity up t $r=12a$, to then decreasing the attraction until $r\approx14.1205a$, where $F^{e}_{MM}=0$. The minimum of the anions RDF is $g^{minimum}_{M-}(r=13.5a)=0.958615$. As a result, we see in Fig.\ref{Fig_1}, the electrical field going from positive values, at $r=5a$, due to the central macroion charge, down to zero, at $r\approx7.97a$, then to a minimum, at $r=12a$, where the cations RDF ($g_{M+}(r)$) goes from its maximum, at $r\approx8.625$, to its bulk value, at $r\approx12.25a$. Then $E_{MM}(r)$ increases up to zero, at $r\approx14.16a$, and then to a maximum, at $r\approx16.25a$, now due to the first layer of macroions charge, located at $r\approx13.5935a$. The first maximum of the entropic force occurs at $r\approx10.92a$, which seems to correspond to a maximum of adsorbed anions plus cations, from $r=5.5a$ to $r\approx11a$, i.e., in Fig.~\ref{Fig_2} we see a strong adsorption of anions, much above its bulk value, from $r=5.5a$, the central-macroion-anion contact value, up to $r\approx10.08$, where they reach their bulk value (i.e., where $g_{M-}(r=10.08a)\approx1$), then down to a minimum, at $r\approx13.5$, whereas the cations go from its lower value, at $r=5.5a$ to their bulk value at $r\approx7.25$, and then to a maximum at $r\approx8.625$, and then again to its bulk value, at $r\approx12.25a$. From its maximum value $F^{s}_{MM}(r)$ decreases down to zero, for $r\approx14.145$. Notice that, although the points of zero entropic and electric force are very close to each other, they do not coincide. The net force, i.e., the mean force, in this region is repulsive, implying that the repulsive entropic force overcomes the attractive electric force. We already pointed out that a point of zero force occurs at $r\approx13.59a$. 

The next maximum of the electrical force, and minimum of the entropic force coincide at $r\approx16.25a$. From the observation of RDFs of the macroions, anions and cations (see Fig.~\ref{Fig_2}), between $r\approx13.59a$ and $r=26.22a$, the positive electrical force is due to the positive charge of the first layer of macroions, while the entropic attraction seems to come from the pressure of the next layer of cations-anions-macroions. The first minimum of the effective force is at $r=15a$, and does not coincide with the nearby minimum of the entropic force, while its second maximum, at $r\approx23.75a$, coincide with the maximum entropic force component. This structure and correlation of the RDFs, Fig.~\ref{Fig_2},  with the forces presented in Fig.~\ref{Fig_3}, are oscillatory $\forall r$. The $\lim_{r\to\infty} F^{j}_{Mi}(r)\to 0$, $\forall j$. 

In between the subsequent layers of macroions, after the location of the second layer of macroions at, $r\approx26.22a$, a similar mechanism seems to occur, i.e., in Fig.~\ref{Fig_2} we see first an adsorption of anions, above their bulk value, followed by an adsorption of cations, which together with the charge of the first layer of macroions, seems to be related to the repulsive, and then attractive force of the electrical component, as a function $r$. Whereas the entropic force goes first from attractive, and then repulsive. So, why the electrical force is attractive?   

\begin{figure}
\resizebox{0.9\columnwidth}{!}{
  \includegraphics{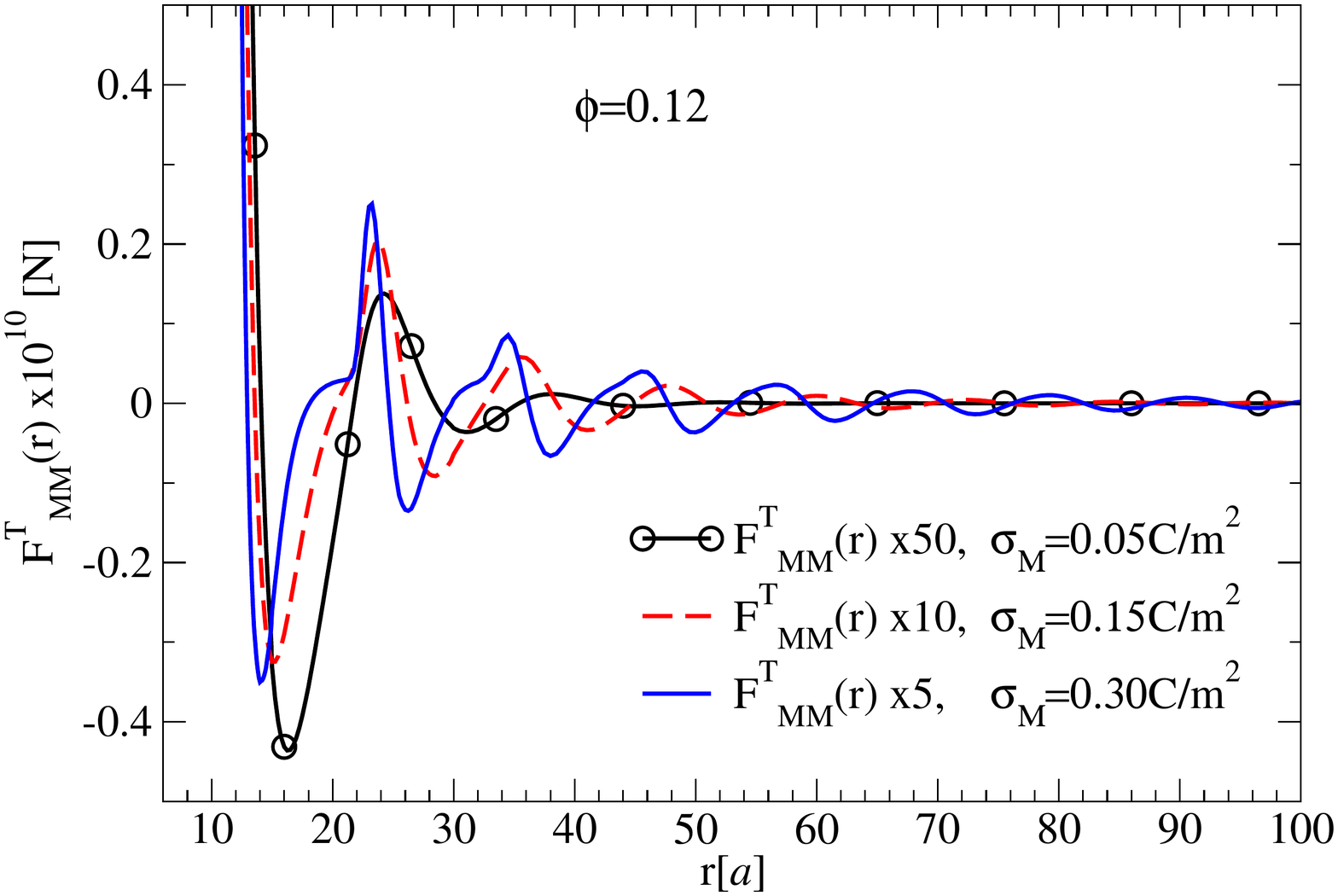} }
\caption{Mean force among like-charged macroions, for different macroions' surface charge densities, corresponding, from lower to higher charge densities, to $Z_M\approx18, 53$, and $106$, respectively. Notice the scaling factors. $r$ is referred to the center of the reference macroion. $\rho_{salt}=0.1$M, $a_M=10a$.}
\label{Fig_4}       % Give a unique label
\end{figure}

To try to understand the role of the macroions charge, and volume fraction, on the total mean force, in Figs.~\ref{Fig_4} we show this force for three different charge densities on the macroions, $\sigma_M=0.05, 0.15$ and $0.30 C/m^2$, which correspond to $Z_M\approx18, 53$, and $103$, respectively. We keep the volume fraction constant, with $\phi=0.12$. The forces are oscillatory. The higher the macroions' charge, the more intense is the force (notice the different scales used in the figure). The higher the charge the closer and intense is the first minima of the force among the like-charged macroions, i.e., the stronger the attraction among like-charged particles. We have performed calculations of the total mean force, for a fixed macroions charge, and three different volume fractions: $\phi=0.06, 0.12$, and $0.24$ (not shown). We find that the higher the volume fraction, the closer the macroions, and stronger, and closer the attractive minima of the total mean force. This is to be expected, since higher volume fraction implies les available volume, and stronger the entropic component of the force. However, it is a bit puzzling why higher charge implies stronger attraction among the macroions. The explanation we think is due to a combination of two effects: a higher adsorption of anions in-between layers of macroions, and that the charge repulsion among like-charged macroions also increases the excluded volume of the system, then increasing the entropic attractive force. This like-charge excluded volume, depends on the amount of charge on the macroparticles and, clearly, of the system's temperature~\cite{deutsch1981}. The shoulder observed around $r=20a$, in the $\sigma_M=0.30 C/m^2$ curve, results from the higher gathering of the macroions, i.e., the higher the macroions charge, the closer they become among them (see Table~\ref{tab:1}). 

As discussed above, beyond the fact that Eqs.~\eqref{trainee} and~\eqref{trainee2} are non-linear integral equations, the mean electrostatic potential, $\psi_{ij}(r)$, and mean potential of the short-range force, $\omega^{sr}_{ij}(r)$, are entangled, and, thus, also are entangled the electric, $F^{e}_{MM}(r)$, and entropic, $F^{s}_{MM}(r)$, components of the total force, $F^{T}_{MM}(r)$. We have performed calculation for a zero charge macroparticles, at, say, $\phi=0.12$, and find that the entropic force (not shown), which in these conditions is equal to the total mean force, is much more less intense and of much shorter range (see Tab~\ref{tab:1}). Therefore, we may conclude that higher macroions charge increases the gathering of the macroion, due to an important adsorption of anions around the macroions, which produces an inversion of the induced charge, $~\sigma_M(r)$, i.e., the electric field, $E_M(r)$, but also due to an an increase of the macroion-macroion electrostatic repulsion, which in turn increases the entropic attraction, i.e., for finite macroions' volume fractions, excluded-volume interactions are accentuated, and like-charged repulsion contributes to an even lower available volume. 

In Table~\ref{tab:1} we display the locations of the first six layers of macroions, as a function of the distance $r/a$ to the central macroion, for different macroions charge. We see that as the charge increases, from $\sigma_M(r)=0.050 C/m^2$ to $\sigma_M(r)=0.300 C/m^2$, the macroion layers compact around the central macroion. For $\sigma_M(r)=0.000 C/m^2$, only the entropic force is present, and the first layer of macroions is in contact with the central macroion ($g_{MM}(r=10a)=1.5636$). For $\sigma_M(r)=0.005 C/m^2$, the first layer of macroions is still in contact with the central macroion, but with a bit less probability ($g_{MM}(r=10a)=1.5634$), and the subsequent layers spread a little bit more. For $\sigma_M(r)=0.025 C/m^2$, the first layer of ions is repelled away from the central macroion, and the subsequent layers of macroions spread further away. Also notice that the macroions charge increases the range of the correlation. For the higher charges display in table~\ref{tab:1}, the number of detected macroions layers go as far as $r~\approx600a$. In short, for uncharged or very little charged macroions the macroion-macroion attraction is due to the entropic force, resultant of considering the macroions volume. Higher volume fractions increases the macroions attraction. At still some very low macroions charge, i.e., $\sigma=0.025 C/m^2$, the first layer of macroions around the central particle is repelled away, as a result of the electrical repulsion. However, for low to high macroions charge, higher charge produces stronger macroions attraction.

\begin{figure}
\resizebox{1.05\columnwidth}{!}{
  \includegraphics{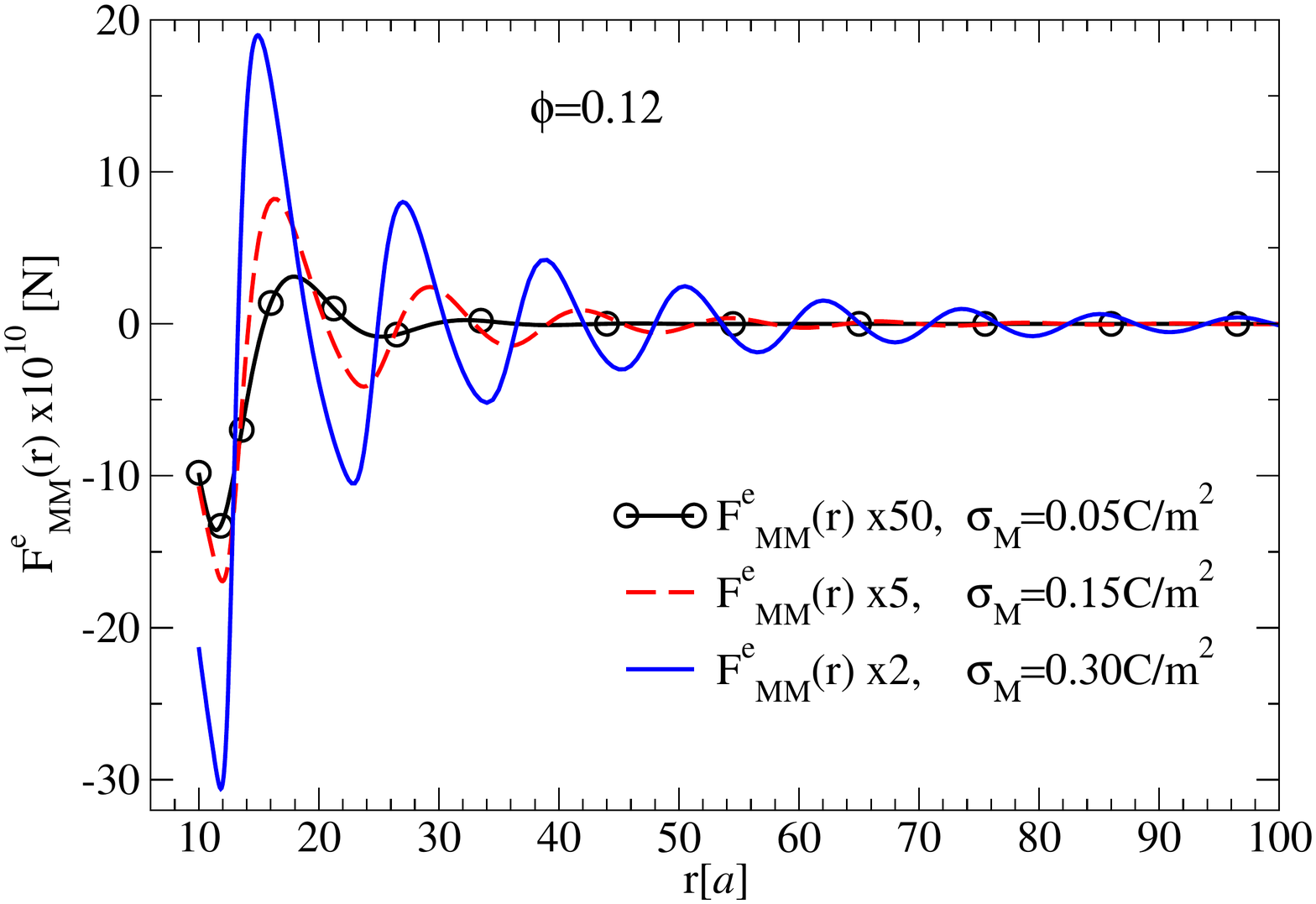} 
  \includegraphics{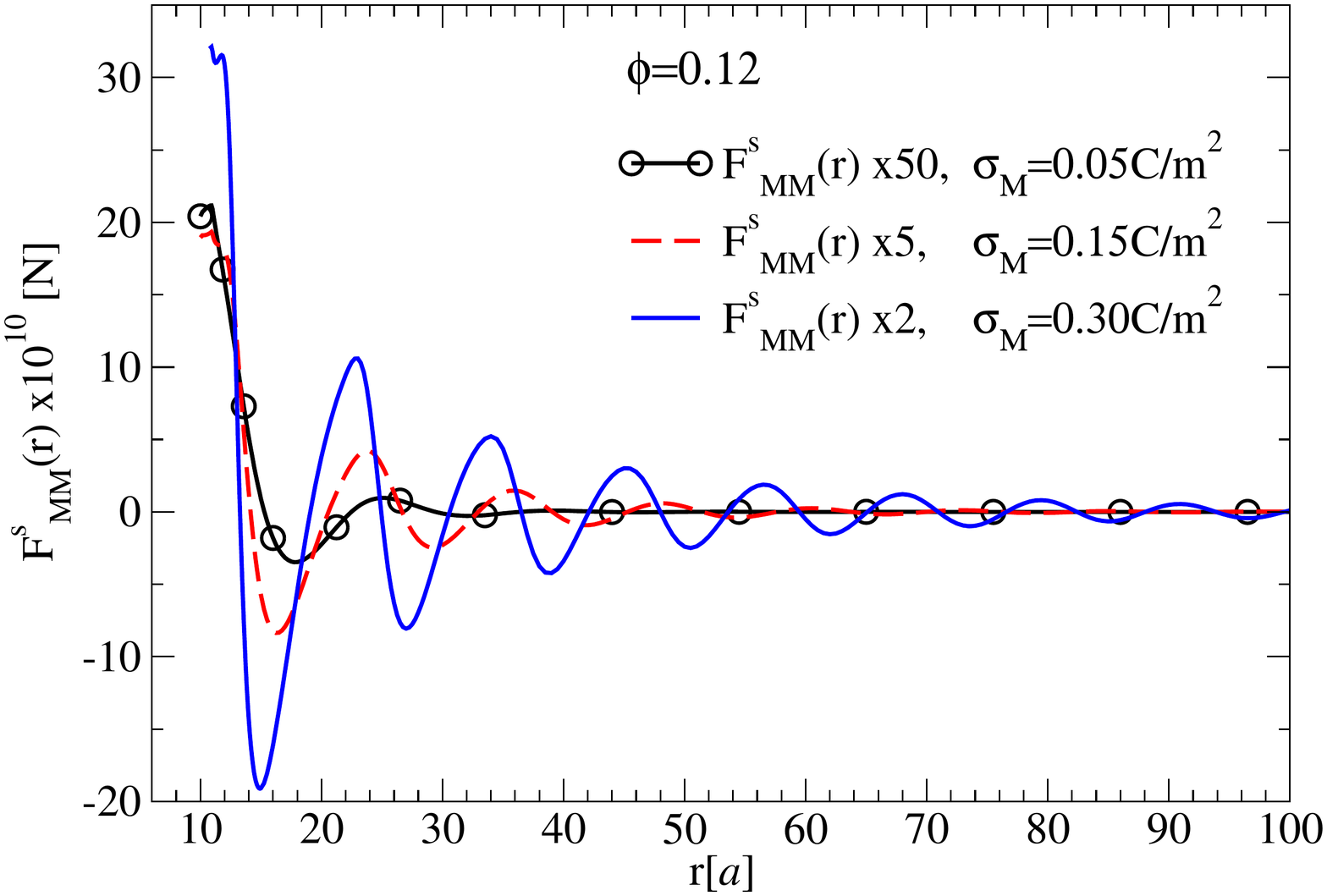}}
\caption{Electrical, $F^{e}_{MM}(r)$, and entropic, $F^{s}_{MM}(r)$, components of the mean force among like-charged macroions, for different macroions' surface charge densities. $r$ is referred to the center of the reference macroion. Notice the scaling factors. $\rho_{salt}=0.1$M, $a_M=10a$. }
\label{Fig_5}       % Give a unique label
\end{figure}

 In Fig.~\ref{Fig_5}, the electric and entropic components of the total force are presented, for the same three different macroions charge, and at a fixed volume fraction, $\phi=0.12$, as in Fig.~\ref{Fig_4}, showing that in both cases their minima are all more intense, and closer to the reference macroion, for higher macroions' charge density. Being the nature of the electrical force among like-charged particles repulsive, a large attractive increase in the intensity, is probably due to the very high adsorption of anions to the positive macroions (see Fig.~\ref{Fig_2}). This attraction is enhanced by the also strong attractive contribution of the entropic force, as a result of the pressure of the external layers of the fluid, to the inner layers of ions. Additionally, one should keep in mind that both, the electric and entropic components of the mean force are entangled and, hence, a simple separation of the two contributions are not independent.
 
The shoulder seen in Fig,~\ref{Fig_4}, is the result of a very large increase in the first maxima of the entropic force components. The higher the charge the longer the macroion-macroion force correlation. This is also true, for higher volume fractions and/or larger macroions size. The behavior of the electric force, as the macroions charge increases show its impact in the decrease of the accesible volume. Both, the electric and force components increase their intensity and correlation, with increasing macroions charge.

\subsection{Long-range attractive and repulsive forces: Comparison with experimental results}
\label{section:6}

In experimental studies of polymer latex suspensions, the range of a long-range attraction among macroions, at finite concentration, is often reported in terms of the macroion-macroion distance of closest approach, $r_{ca}$, which we define here as the location of the first layer of macroions, around the central macroion, or the first minimum of the macroion-macroion potential of mean force, or the first point of zero total force in Figs.~\ref{Fig_3} to Fig.~\ref{Fig_5}. We have calculated the macroion-macroion radial distribution function for four different macroions diameters ($a_M=10a, 20a, 30a$ and $40a$), at fixed $\phi=0.12$, $\sigma_M=5.355x10^-2$, $\rho_salt=0.1M$ (not shown). Their corresponding distances of macroion-macroion closest approach are $r_{ca}=1.41a_M, 1.33a_M, 1.27a_M$, and $1.24a_M$, respectively. As a general behavior, larger macroions size implies higher concentration. of the first layer of the attracted macroions, and shorter values of $r_{ca}$, although of course, in units of $a$, $r_{ca}$ increases with increasing $a_M$.

\begin{figure}
\resizebox{0.9\columnwidth}{!}{
  \includegraphics{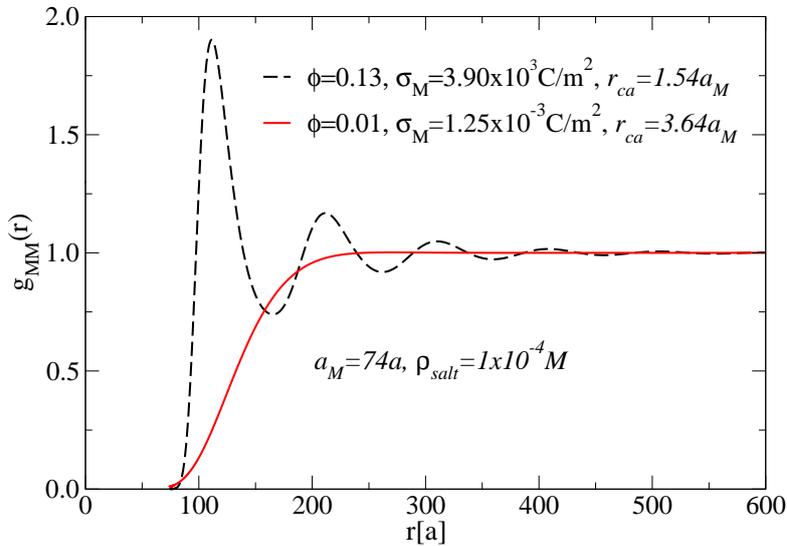} }
\caption{Radial distribution function, around the central macroion, for two different volume fractions and surface charge densities. $r$ is referred to the center of the corresponding reference macroion.}
\label{Fig_7}       % Give a unique label
\end{figure}

A \textit{direct} comparison of our theory with experimental for very large macroions, such as those reported for polystyrene latex particles~\cite{ise83,asher1991,ise1994,tata2008}, where the macroion sizes are in the range of 3000{\AA}  to 8000{\AA} is beyond the scope of this paper. Fortunately, however, there are some experimental data for smaller polystyrene latex particles, such as those reported por R. H. Ottewill~\cite{ottewill1989}. In Fig.~\ref{Fig_7} we presente a comparison of the radial distribution function for two different sets of parameters: (a) $\phi=0.01$, $\sigma_M=1.25x10^{-3}C/m^2$, and (b) $\phi=0.13$, $\sigma_M=3.90x10^{-3}C/m^2$. In both cases $a_M=74a=314.5$ {\AA}, and $\rho_{salt}=1.0x10^-4M$. The experimental data parameters are (a) $\phi=0.01$ and (b) $\phi=0.13$. In both cases $a_{Mexp}\approx310${\AA}$\approx73a_{Mexp}$ and $\rho_{salt}=1.0x10^-4M$. He did not report measured macroions' surface charge densities, but fitted macroions' surface mean electrostatic potentials, of the order 50mV. In our calculations, the macroions's surface potential are both of the order of 17 mV. Hence, although we can not make a direct quantitative comparison with these experimental results, our calculations parameters closely resemble those reported by Ottewill. The qualitative agreement of our Fig.~\ref{Fig_7} with the figure 4 of Ottewill's paper is very good. Moreover, he reported values of $r_{ca}$ of 1000{\AA}=3.22$a_{Mexp}$ and 500{\AA}=1.61$a_{Mexp}$, obtained from his radial distribution functions for his $\phi=0.01$ and $\phi=0.13$, respectively. In our case, we find $r_{ca}$ of 1145.8{\AA}=3.64$a_{M}$ and 483{\AA}=1.54$a_{Mexp}$, for $\phi=0.01$ and $\phi=0.13$, respectively.  

We did perform calculations for macroions as large as $340a_M$, for different salt concentrations, macroions's charge and volume fractions (not shown), and as a general trend we find that the $r_{ca}$ parameter scales with the macroion's size, and that increasing $a_M$ and/or $\phi$ and/or $\sigma_M$ and/or $\rho_{salt}$, decreases $r_{ca}$. Nevertheless, different combinations of these parameters can produce very large values of $r_{ca}$, as we have shown. Clearly, the larger the macroion's size, the larger the location of the first attractive minimum in the macroion-macroion potential of mean force. Experimental results for macroions of 6000{\AA} of diameter, and surfaces charges of $\sigma=2.7x10^{-2} C/m^2$, and $\sigma=3.0x10^{-3}C/m^2$, reported by Tata et al.~\cite{tata2008}, have values of $r_{ca}$=2.42$a_{Mexp}$ and $r_{ca}$=2.47$a_{Mexp}$, respectively. Those reported by Ise et al.~\cite{ise83} show similar results for their 4190{\AA} diameter, and $\sigma=7.25x10^-2C/m^2$, polystyrene particles, i.e., for $\phi=0.0075$ and $\phi=0.112$, they find $r_{ca}= 3a_{mexp}$ and $r_{Mexp}=1.695a_{Mexp}$.

While our $r_{ca}$ results confirm the existence of a long-range macroions attraction, and have a good qualitative agreement with some experimental data, we wish to emphasize the relevance of the truly very long range of the oscillatory macroion-macroion radial distribution function and force, which may have relevant consequences in several soft condensed matter and biological systems~\cite{alfredo2017,alfredo2018}.

%%%%%%%%%%%%%%%%%%%%%%%%%%%%%%%%%%%%%%%%%%%%%%%%%%%%%%%%%%%%%%%%%%
\section{Discussion}
\label{sec:7}

The objective of this research was to explore the nature of the mean forces present in a highly charged macroions solution, at a finite volume fraction, $\phi$, with a well established liquid theory. The experiments of N. Ise et al., where an order-disorder coexistence regions are reported~\cite{ise83,ise1993,ise94,tata1997}, imply the existence of long-range attractive forces among like-charged macroions. However, these long-range forces can not be explained in terms of conventional direct interaction potentials, which are of short range~\cite{israelachviliNature08}. Semi-phenomenological long-range potentials have been proposed in the past~\cite{sogami84,ise96,sogami2012}. N. Ise et al.~\cite{ise66,sogami84,ise2010} have proposed that this attraction is due to ``\textit{an intermacroion attraction through the intermediary of counterions balanced by intermacroion repulsion}''. While in this article we did not pretend to find a coexistence curve, and do not take a strong position on the details of the experimental results, we solved the well-established HNC/MSA liquid theory, applied to a colloidal primitive model, obtained the macroion-macroion, macroion-anion, and macroion-cation radial distribution functions, and calculated the electrical field, and the mean forces among the like-charged macroions. Our study shows that higher macroions charge increases the gathering of the macroion, due to an important adsorption of anions around the macroions, which produces an inversion of the induced charge, $~\sigma_M(r)$, i.e., the electric field, $E_M(r)$, but also due to an an increase of the macroion-macroion electrostatic repulsion, which in turn increases the entropic attraction, i.e., for finite macroions' volume fractions, excluded-volume interactions are accentuated, and like-charged repulsion contributes to an even lower available volume. 

Our findings are in agreement with two aspects of the experimental results of B. V. R. Tata et al.~\cite{tata1994,tata1997,ise2000,tata2008}: like charged particles get closer among them with higher electrical charge, and there is a very long-range attractive-repulsive correlation.  ~\textit{Our calculations are with no assumption of a long-range attractive potential or confining wall, and the system is certainly at equilibrium, since the integral equations are basically a chemical potential balance equation}. While our results seem to indicate that the high adsorption of anions in between two neighboring layers of macroions, explain in part the attraction among the like-charged macroions, as proposed by Ise et al.~\cite{ise66,sogami84,ise2010}, the fact that the electrical and entropic forces are entangled, through the non-linearity of the radial distribution functions equations of the colloidal primitive model (Eqs.~\eqref{trainee} and ~\eqref{trainee2}), imply that the electrical component also contributes to an excluded volume effect, and, hence, is also in part responsible for the attractive force among like-charged macroions.

In particular, our radial distribution function calculations, for macroions' parameters corresponding to those of the experimental radial distribution functions reported by R. H. Ottewill~\cite{ottewill1989} show a very good qualitative agreement. In particular, in relation to the range of the macroions long-range attractive interaction. 

As a general result we find that the range of the attractive-repulsive forces, among the macroions, scales with the macroion's size, and that increasing the macroion size and/or their volume fraction and/or charge and/or the salt concentration, increases the macroions attraction. Nevertheless, different combinations of these parameters can produce very long-range macroion-macroion attractions, as we have shown. Clearly, the larger the macroion's size, the larger the range of the first attractive minimum in the macroion-macroion potential of mean force.

As a main conclusion we stress that our theoretical calculations are based on a well-stablished integral equation theory, shown in the past to be in agreement with computer simulation data, we have not used any adjustable parameter, and have proved the existence of very long-range attractive-repulsive forces, in macroions dispersions at finite volume fraction, and are in agreement with experimental results.

We gratefully acknowledge the support of CONACyT, M\'exico, through the Project 169125.

%\bibliography{Long_Ranged_Forces_EPJ_ST}

\begin{thebibliography}{10}
\providecommand{\url}[1]{{#1}}
\providecommand{\urlprefix}{URL }
\expandafter\ifx\csname urlstyle\endcsname\relax
  \providecommand{\doi}[1]{DOI \discretionary{}{}{}#1}\else
  \providecommand{\doi}{DOI \discretionary{}{}{}\begingroup
  \urlstyle{rm}\Url}\fi

\bibitem{pieransky83}
P.~Pieransky, Contemp. Phys. \textbf{24}, 25 (1983)

\bibitem{kamenetzky94}
E.A. Kamenetzky, L.G. Magliocco, H.P. Panzer, Science \textbf{263}, 207 (1994)

\bibitem{wennerstrom}
D.F. Evans, H.~Wennerstr{\"o}m, \emph{The Colloidal Domain: Where Physics,
  Chemistry, Biology, and Technology Meet}, 2nd edn. (Wiley-VCH, New York,
  1999)

\bibitem{vatamanu}
J.~Vatamanu, D.~Bedrov, J. Phys. Chem. Lett. \textbf{6}, 3594 (2015)

\bibitem{clark1979}
N.A. Clark, A.J. Hurd, B.J. Ackerson, Nature \textbf{281}, 57 (1979)

\bibitem{ise80}
N.~Ise, T.~Okubo, Acc. Chem. Res. \textbf{13}, 303 (1980)

\bibitem{ottewill1989}
R.H. Ottewill, Langmuir \textbf{5}, 4 (1989)

\bibitem{ise83}
N.~Ise, T.~Okubo, M.~Sugimura, K.~Ito, H.J. Nolte, J. chem. Phys \textbf{78},
  536 (1983)

\bibitem{ise1993}
H.~Matsuoka, H.~Nakamura, H.~Okumura, T.~Ono, I.S. Sogami, J.Y. Ueno,
  H.~Yoshida, T.~Yoshiyamal, Langmuir \textbf{9}, 394 (1993)

\bibitem{ise94}
K.~Ito, H.~Yoshida, N.~Ise, Science \textbf{263}, 66 (1994)

\bibitem{tata1997}
B.V.R. Tata, E.~Yamahara, P.V. Rajamani, N.~Ise, Phys. Rev. Letts. \textbf{78},
  2660 (1997)

\bibitem{israelachviliNature08}
Y.~Min, M.~Akbulut, K.~Kristiansen, Y.~Golan, J.~Israelachvili, Nat. Mater.
  \textbf{7}(7), 527 (2008)

\bibitem{sogami84}
I.~Sogami, N.~Ise, J. Chem. Phys. \textbf{81}, 6320 (1984)

\bibitem{ise96}
N.~Ise, Ber. Bunsenges. Phys. Chem. \textbf{10}, 841 (1996)

\bibitem{sogami2012}
I.S. Sogami, Chem. Lett. \textbf{41}, 1331 (2012)

\bibitem{palberg1994}
M.~Evers, N.~Garbow, D.~Hessinger, T.~Palberg, Phys. Rev. Letts. \textbf{72},
  786 (1994)

\bibitem{grier1996}
J.C. Crocker, D.G. Grier, Phys. Rev. Letts. \textbf{77}, 1897 (1996)

\bibitem{hansen97}
R.~van Roij, J.P. Hansen, Phys. Rev. Lett. \textbf{79}, 3082 (1997)

\bibitem{schmitz02}
K.S. Schmitz, Phys. Rev. E \textbf{65}, 061402(1) (2002)

\bibitem{tata1994}
B.V.R. Tata, A.K. Arora, Phys. Rev. Letts. \textbf{72}, 787 (1994)

\bibitem{ise2000}
B.V.R. Tata, N.~Ise, Phys. Rev. E \textbf{61}, 983 (2000)

\bibitem{tata2008}
B.V.R. Tata, P.S. Mohanty, M.C. Valsakumara, Solid State Communications
  \textbf{147}, 360 (2008).
\newblock And references therein

\bibitem{ise66}
N.~Ise, T.~Okubo, J. Phys. Chem. \textbf{70}, 1930 (1966)

\bibitem{ise2010}
N.~Ise, Phys. Chem. Chem. Phys. \textbf{12}, 10279 (2010)

\bibitem{verwey48}
E.J.W. Verwey, J.T.G. Overbeek, \emph{Theory of the Stability of Lyophobic
  Colloids} (Dover, New York, 1948)

\bibitem{stillinger1960}
F.H. Stillinger, J.G. Kirkwood, J. Chem. Phys. \textbf{33}, 1282 (1960)

\bibitem{mcquarrie80}
M.~Medina-Noyola, D.~McQuarrie, J. Chem. Phys. \textbf{73}, 6279 (1980)

\bibitem{schmitz93}
K.S. Schmitz, \emph{Macroions in Solution and Colloidal Suspensions} (VCH
  Publishers, New York, 1993)

\bibitem{Kjellander86}
R.~Kjellander, S.~Mar\v{c}elja, Chem. Phys. Lett. \textbf{127}, 402 (1986)

\bibitem{levin99}
A.~Diehl, M.N. Tamashiro, M.C. Barbosa, Y.~Levin, Physica A \textbf{274}, 433
  (1999)

\bibitem{belloni2000}
L.~Belloni, J. Phys.: Condens. Matter \textbf{12}, R549 (2000)

\bibitem{levin2016}
T.~Colla, M.~Girotto, A.P. dos Santos, Y.~Levin, J. Chem. Phys. \textbf{145},
  094704 (2016)

\bibitem{belloni1998}
L.~Belloni, Colloids Surfaces A: Physicochem. Eng. Aspects \textbf{140}, 227
  (1998)

\bibitem{bocquet2002}
L.~Bocquet, E.~Trizac, M.~Aubouy, J. Chem. Phys. \textbf{117}, 8138 (2002)

\bibitem{ise99}
N.~Ise, T.~Konishi, B.V.R. Tata, Langmuir \textbf{15}, 4176 (1999).
\newblock And references therein

\bibitem{medinanoyola2011}
L.E. S·nchez-DÌaz, G.A. Mendez-Maldonado, M.~Gonz·lez-Melchor, H.~Ruiz-Estrada,
  M.~Medina-Noyola, J. Chem. Phys. \textbf{135}, 014504 (2011)

\bibitem{linse99}
P.~Linse, V.~Lobaskin, Phys. Rev. Lett. \textbf{83}, 4208 (1999)

\bibitem{linse00}
P.~Linse, V.~Lobaskin, J. Chem. Phys. \textbf{112}, 3917 (2000)

\bibitem{lobaskin03}
V.~Lobaskin, K.~Qamhieh, J. Phys. Chem. B \textbf{107}, 8022 (2003)

\bibitem{poirier1954}
J.G. Kirkwood, J.C. Poirier, J. Phys. Chem. \textbf{591-596}, 1282 (1954)

\bibitem{McQuarrie}
D.A. McQuarrie, \emph{Statistical Mechanics} (Harper and Row, New York, 1976)

\bibitem{Lozada-Cassou92a}
M.~Lozada-Cassou, in \emph{Fundamentals of Inhomogeneous Fluds}, ed. by
  D.~Henderson (Marcel Dekker, New York, 1992), chap.~8

\bibitem{attard96}
P.~Attard, in \emph{Advances in Chemical Physics}, vol. XCII, ed. by
  I.~Prigogine, S.A. Rice (John Wiley and Sons, Inc., New York, 1996)

\bibitem{carnie1981}
S.L. Carnie, D.Y.C. Chan, D.J. Mitchell, B.W. Ninham, J. Chem. Phys.
  \textbf{74}, 1472 (1981)

\bibitem{manzanilla2011a}
H.M. Manzanilla-Granados, F.~Jim{\'e}nez-{\'A}ngeles, M.~Lozada-Cassou,
  Colloids and Surfaces A: Physicochem. Eng. Aspects \textbf{376}, 59 (2011)

\bibitem{manzanilla2011b}
H.M. Manzanilla-Granados, F.~Jim{\'e}nez-{\'A}ngeles, M.~Lozada-Cassou, J.
  Phys. Chem. B \textbf{115}, 12094 (2011)

\bibitem{manzanilla2013}
H.~Manzanilla-Granados, M.~Lozada-Cassou, J. Phys. Chem. B \textbf{117}, 11812
  (2013)

\bibitem{kjellander1992}
H.~Greberg, R.~Kjellander, Chem. Phys. Letts. \textbf{200}, 76 (1992)

\bibitem{zuckerman2001}
D.M. Zuckerman, M.E. Fisher, S.~Bekiranov, Phys. Rev. E \textbf{64}, 011206
  (2001)

\bibitem{odriozola2017}
G.~Odriozola, M.~Lozada-Cassou, Fortschr. Phys. 65, No. 6?8, 1600072
  \textbf{65}, 1600072 (2017)

\bibitem{Friedmanbook}
H.L. Friedman, \emph{A course in Statistical Mechanics} (Prentice-Hall, Inc.,
  1985)

\bibitem{kazuo77}
K.~Hiroike, Mol. Phys. \textbf{33}, 1195 (1977)

\bibitem{sanchez92}
J.E. S{\'a}nchez, M.~Lozada-Cassou, Chem. Phys.Lett. \textbf{190}(3,4), 202
  (1992)

\bibitem{Yeomans1}
L.~Yeomans, E.S.S.E. Feller, M.~Lozada-Cassou, J. Chem. Phys. \textbf{98}, 1436
  (1993)

\bibitem{lawless}
W.F. Lawless, Journal of Mathematical Psychology \textbf{78}, 51 (2017)

\bibitem{alfredo2017}
J.A. Gonz{\'a}lez-Calder{\'o}n, M.~Ch{\'a}vez-P{\'a}ez, E.~Gonz{'a}lez-Tovar,
  M.~Lozada-Cassou, cond-mat.soft \textbf{arXiv:1712.00467v1}, 1 (2017)

\bibitem{alfredo2018}
J.A. Gonz{\'a}lez-Calder{\'o}n, M.~Ch{\'a}vez-P{\'a}ez, E.~Gonz{'a}lez-Tovar,
  M.~Lozada-Cassou, J. Phys. Chem. B \textbf{122}, 7002 (2018)

\bibitem{henderson1983}
D.~Henderson, M.~Lozada-Cassou, L.~Blum, Journal of Chem. Phys. \textbf{79},
  3055 (1983)

\bibitem{LozadaJCP82}
M.~Lozada-Cassou, R.~Saavedra-Barrera, D.~Henderson, J. Chem. Phys.
  \textbf{77}, 5150 (1982)

\bibitem{gonzalez85}
E.~Gonz{\'a}lez-Tovar, M.~Lozada-Cassou, D.~Henderson, J. Chem. Phys.
  \textbf{83}, 361 (1985)

\bibitem{gonzalez89}
E.~Gonz{\'a}lez-Tovar, M.~Lozada-Cassou, J. Chem. Phys. \textbf{93}, 3761
  (1989)

\bibitem{LozadaJCP90a}
M.~Lozada-Cassou, E.~D{\'i}az-Herrera, J. Chem. Phys. \textbf{92}, 1194 (1990)

\bibitem{jimenez04}
F.~Jim{\'e}nez-{\'A}ngeles, M.~Lozada-Cassou, J. Phys. Chem B. \textbf{108},
  7286 (2004)

\bibitem{odriozola2009}
F.~Jim{\'e}nez-{\'A}ngeles, G.~Odriozola, M.~Lozada-Cassou, J. Phys.: Condens.
  Matter \textbf{21}, 424107 (2009)

\bibitem{deutsch1981}
C.~Deutsch, Y.~Furutani, M.~Gomber, Phys. Rep. \textbf{69}(2), 85 (1981)

\bibitem{ise1994}
N.~Ise, M.V. Smalley, Phys. Rev. B \textbf{50}, 16722 (1994)

\bibitem{asher1991}
P.A. Rundquist, S.~Jagannathan, R.~Kesavamoorthy, C.~Brnardic, S.~Xu, S.A.
  Asher, J. Chem. Phys. \textbf{94}, 711 (1991)

\end{thebibliography}
%\bibliographystyle{spphys}

%
%\begin{thebibliography}{}
%Author, Journal \textbf{Volume}, (year) page numbers
%\bibitem{Lozada1}
%M. Lozada-Cassou, J. Chem. Phys. \textbf{1}.(2018) 12-13
%Author, \textit{Book title} (Publisher, place year) page numbers
% etc
%\end{thebibliography}
\end{document}